\documentclass[prl,aps,onecolumn,superscriptaddress,preprint]{revtex4}
\usepackage[dvips,final]{graphicx}
\usepackage{txfonts}

\newcommand{\Mdot}{{\dot M}}

\begin{document}
\title{Luminosity, redshift and gas abundance in general relativistic radiation
hydrodynamics}
\author{Janusz Karkowski}
\affiliation{M. Smoluchowski Institute of Physics, Jagiellonian University,
Reymonta 4, 30-059 Krak\'{o}w, Poland}
\author{Edward Malec}
\affiliation{M. Smoluchowski Institute of Physics, Jagiellonian University,
Reymonta 4, 30-059 Krak\'{o}w, Poland}
\author{Krzysztof Roszkowski}
\affiliation{M. Smoluchowski Institute of Physics, Jagiellonian University,
Reymonta 4, 30-059 Krak\'{o}w, Poland}
\author{Zdobys\l aw \'Swierczy\'nski}
\affiliation{Pedagogical University, Podchor\c a\.zych 1, Krak\'{o}w, Poland}

\begin{abstract}
Quasi-stationary flows of  gas accreting onto a compact center are analyzed
in the framework of general-relativistic radiation hydrodynamics, under
assumptions of spherical symmetry and thin gas approximation. Numerical 
investigation shows that luminosity, redshift and gas abundance are correlated. 
The  gas  can constitute up to one third of the total mass of brightest low-redshift 
sources, but its abundance goes down to $1/30$ for  sources with luminosities close to
the Eddington limit.   
 
\end{abstract}

\maketitle

We investigate a steady gas accretion onto a compact core in  the framework
of general relativity.   The main goal of this letter 
is to show that bright sources --- with the luminosity approaching the Eddington limit 
---  must contain  a significant fraction of gas.
Our model assumes  spherical symmetry, polytropic equation of state and  thin gas
approximation in the transport equation   \cite{Mihalas}. 
  
 We use comoving coordinates $t, r, 0\le \theta \le \pi , 0\le \phi < 2\pi$:
time, coordinate radius and two angle variables, respectively. The metric is
\begin{equation}
ds^2=-N^2dt^2+\hat adr^2 +R^2d\Omega^2
\label{1}
\end{equation}
where $R$ denotes the area radius. The radial velocity of gas is given by
$U = \frac{1}{N} \frac{dR}{dt}$.  
 
The energy-momentum tensor reads $T_{\mu \nu }=T_{\mu \nu }^B +T_{\mu \nu }^E$,
where the baryonic part is given by $T_{\mu \nu }^B =
(\rho +p )U_\mu U_\nu +pg_{\mu \nu }$ with the time-like and normalized
four-velocity $U_\mu $, $U_\mu U^\mu =-1$.
The radiation part has only four nonzero components: $T_0^{0E}\equiv -\rho^E=
-T_r^{rE}$ and $T^E_{r0}= T^E_{0r}$.
A comoving observer would measure local mass densities, material
$\rho =T^{B\mu \nu }U_\mu U_\nu $ and radiation $\rho^E$, respectively. The
baryonic current is defined as $j^\mu \equiv \rho_0 U^\mu$, where $\rho_0$ is
the baryonic mass density. Define $n_\mu$ as the unit normal to a centered
(coordinate) sphere lying in the hypersurface $t=const$ and $k$ as the related
mean curvature scalar, $k={R\over 2}\nabla_i n^i=\frac{1}{\sqrt{\hat a}}\partial_rR$.
The comoving radiation flux density reads
$j = U_\mu n^\nu NT^{\mu E}_\nu /\sqrt{\hat a} = NT^{0E}_r /\sqrt{\hat a}$.
The baryonic matter satisfies the polytropic equation of state $p=K\rho_0^\Gamma $ 
(with constants $K$ and $\Gamma $). The  internal energy $h$ and the rest and baryonic  mass
densities are related by $\rho =\rho_0+h$, where   $h=p/(\Gamma -1)$.
  
The equation  
\begin{equation}
\nabla_\mu j^\mu = 0
\label{7}
\end{equation}
expresses the conservation of baryonic matter.

There are four conservation equations resulting (due to the contracted Bianchi identities)
from the Einstein equations, namely $\nabla_\mu T^{\mu B}_\nu =
-\nabla_\mu T^{\mu E}_\nu =F_\nu$ (here $\nu = 0, r$). The quantity $F_\nu $ is
the radiation force density and it describes interaction between baryons
and radiation. The present formulation of general-relativistic radiation hydrodynamics
agrees with that of Park \cite{Park}, Miller and Rezzola
\cite{Rezzola} and (on a Schwarzschildean background) Thorne
\textsl{et.\ al} \cite{ThorneF}. 

One can solve formally the Einstein  constraint equations $G_{\mu 0}=8\pi T_{\mu 0}$,
arriving at (\cite{Iriondo},\cite{EM99})
\begin{eqnarray}
&& k = \sqrt{1-\frac{2m(R)}{R} +U^2}.
\label{8}
\end{eqnarray}
Above   $m(R)$ is the quasilocal mass given by
\begin{equation}
m(R)=m-4\pi \int_R^{R_\infty }dr r^2\left( \rho +\rho^E + \frac{Uj}{k}\right) .
\label{9}
\end{equation}
The integration in (\ref{9}) extends from $R$ to the outer boundary $R_\infty$.
 A ball of gas is  comprised between a hard core of a radius $R_0$ and 
sphere $S_\infty$ of a radius $R_\infty$. Its external boundary is connected 
to the Schwarzschild vacuum spacetime by a transient zone of a negligible (due to
special transitory data)  mass. Thence the asymptotic mass
$M$ is approximately equal to $m\left( R_\infty \right)$. Similar picture emerges
in the recent construction of quasistars \cite{Begelman}.

In an alternative, polar gauge foliation, one has a new time $t_S(t,r)$
with $\partial_{t_S}=\partial_t-NU\partial_R$. The expression
$4\pi Nk R^2\left( j\left( 1+ \left( \frac{U}{k}\right)^2\right) +2U\rho^E/k
\right)$ represents the radiation flux measured by an observer located at $R$
in coordinates $(t_S,R)$. 
One can show that 
\begin{equation}
\partial_{t_S} m(R)=
-4 \pi \left( Nk R^2\left( j\left( 1+ \left( \frac{U}{k}\right)^2\right)
+ 2\rho^E{U\over k} \right) +
NU R^2 \left( \rho +p \right) \right)_{R}^{R_\infty}.
\label{10}
\end{equation}
The mass contained in the annulus $(R, R_\infty )$ changes if the fluxes on the right
hand side, one directed outward and the other inward, do not cancel.  

The local baryonic flux will be denoted as $\dot M=-4\pi   UR^2\rho_0$
and its boundary value reads $\dot M_\infty $.
The accretion process is said to be  quasistationary if all relevant
observables  measured at $R$ are approximately constant during time
intervals much smaller than the  runaway instability time scale  $T=M/\dot M_\infty $.  
Analytically, we assume that $\partial_{t_S}X\equiv (\partial_t -
NU\partial_R) X = 0$ for $X=\rho_0$, $\rho$, $j$, $U$\ldots
 
The above assumptions imply,  in the thin gas approximation
\cite{Mihalas}, that $F_0=0$ and the  radiation force density has only one nonzero 
component  $F_r=\kappa kN \rho_0 j$.
The only direct interaction between baryons and radiation is through elastic
Thomson scattering. $\kappa $ is a material constant, depending in particular
on the Thomson cross section $\sigma $, $\kappa =\sigma /\left( 4\pi m_p c
\right)$. $c$ is the speed of light and $m_p$ is the proton mass.

The full system of equations in a form suitable for numerics has been derived elsewhere
\cite{Karkowski2008}. It consists of:

  \begin{enumerate}
\renewcommand{\theenumi}{\roman{enumi}}
\renewcommand{\labelenumi}{\theenumi )}
\item The total energy conservation
\begin{equation}
\Mdot N \frac{\Gamma-1}{\Gamma-1-a^2}+ 2 \Mdot N \frac{\rho^E}{\rho_0} =
4\pi R^2 j N k \left( 1+ \frac{U^2}{k^2}\right) +C ;
\label{19}
\end{equation}
 The constant $C$ is the asymptotic energy flux 
flowing through the sphere of a radius $R_\infty $ (see   (\ref{10})).

\item The local  radiation energy conservation (below $a=\sqrt{dp\over d\rho }$ is the speed of sound)
\begin{eqnarray}
\lefteqn{\left( 1- \frac{2m(R)}{R} \right) \frac{N}{R^2} \frac{d}{dR}
\left( R^2\rho^E\right)=-\kappa kN j\rho_0 + 
2N\left( U \rho^E -kj\right) \frac{dU}{dR} +} \nonumber\\
&& 2k\left( jU -k\rho^E\right) \frac{dN}{dR} + 8\pi NRk
\left( j^2 - j \rho^E \frac{U}{k^2}\right) .
\label{21}
\end{eqnarray}

\item  The relativistic Euler equation    

\begin{eqnarray}
&&\frac{d}{dR} \ln a^2 = -\frac{\Gamma-1-a^2}{a^2- \frac{U^2}{k^2}}
\times \nonumber\\
&&\Biggl[ \frac{1}{k^2 R} \left( \frac{m(R)}{R} -2U^2 + 4\pi R^2
\left( \rho + p + j \frac{U}{k} \right) \right) - \nonumber\\
&&\kappa j \left( 1- \frac{a^2}{\Gamma-1} \right) \Biggr] .
\label{23}
\end{eqnarray}
\item The baryonic mass conservation
\begin{equation}
\frac{dU}{dR} = -\frac{U}{\Gamma -1 -a^2} \frac{d}{dR} \ln a^2 - \frac{2U}{R} +
\frac{4\pi R j}{k}.
\label{24}
\end{equation}
\item The equation for the lapse
\begin{equation}
\frac{dN}{dR}=N \left( \kappa j \frac{\Gamma-1-a^2}{\Gamma-1} + \frac{d}{dR}
\ln \left( \Gamma -1 -a^2 \right) \right) .
\label{25}
\end{equation}
\end{enumerate}

Equations (\ref{19})---\ref{25}) constitute, with $k$
and $m(R)$ given by (\ref{8}) and (\ref{9}), the complete model used in
numerical calculations.  
The asymptotic data are such that $a^2_\infty \gg M/R_\infty \gg U^2_\infty$, which
guarantees the fulfillment of the Jeans criterion for the stability
(see a discussion in \cite{AA} and studies of stability of accreting flows in
newtonian hydrodynamics \cite{Mach}), suggesting in turn the stability of solutions.
One can   put $j_\infty =\rho^E_\infty$. The total luminosity
is well approximated   by $L_0=4\pi R^2_\infty j_\infty $ and it should be related
to the accretion rate by the formula
\begin{equation}
L_0= \alpha \Mdot_\infty  \equiv \left(  1- \frac{N\left( R_0\right)}{k\left( R_0\right)}
\sqrt{1 - \frac{2m\left( R_0\right)}{R_0}} \right) \Mdot_\infty .
\label{26}
\end{equation}
 The last  formula is justified by two arguments. i) In the nonrelativistic limit
one gets $\alpha =|\phi \left( R_0\right) |$, where $\phi $ is the newtonian potential.
 Eq. (\ref{26}) states now  that the   binding energy is transformed into radiation with
the implicit assumption that the heat capacity of the core is negligible. 
ii) The condition of stationarity implies the existence of the approximate time-like 
Killing vector and it appears that $\alpha$ gives the standard measure of the gravitational 
redshift.   If stationary observers detect $\omega_0$ at $R_0$ and $\omega$ at
infinity, and $1/\omega \ll 2M$ (the geometric optics condition --- see
\cite{Karkowski2003} for a discussion) then $\omega = \left( 1-\alpha \right)
\omega_0$. Thus $\alpha$ can be regarded as a proper binding energy and again one arrives 
at formula (\ref{26}).

Let us remark that from above definitions and the equation (\ref{23}) one infers
$L_0 \le 4\pi M/\kappa $, for accretion solutions; it is notable here that the 
limiting luminosity involves the total mass $M$ instead of the mass $m(R_0)$ 
of the central core.

The triple of independent boundary data can consist of  $\alpha$,  $L_0$ and $a^2_\infty$.
These quantities can be determined from observations of highest redshift, total luminosity and asymptotic 
temperature, respectively.  Then one chooses  $j_\infty = \rho^E_\infty =
L_0/\left( 4\pi R^2_\infty \right)$, and the mass accretion rate
$\Mdot = L_0/\alpha$. These data specify supersonic flows up to,
possibly, a bifurcation \cite{Karkowski2008}. In the
case of subsonic flows another boundary condition is needed, for instance the
asymptotic baryonic mass density $\rho_\infty$.  

Eqs.\   (\ref{21}-\ref{25}) are in the  evolution form. Numerical calculation starts from the outer 
boundary $R_\infty$, taking into account Eq. (\ref{19}), and evolves inwards until the equality
$\alpha = 1- \frac{N\left( R\right)}{k\left( R\right) }
\sqrt{1-\frac{2m\left( R\right)}{R}} $ is met at some $R$, denoted as $R_0$ and being 
regarded as the radius of the compact core.  
The numerical integration  employs the 8th order Runge-Kutta method \cite{RK8}. 
Choosing $\rho_{0\infty }$ at random one either obtains 
no solution at all or a subsonic solution. Using the bisection method and automating
the search process, one can obtain a boundary of the
solution set (on the plane $L_0$ -- $ \rho_{0\infty }$), which (interestingly enough)
 appears to bifurcate  from  a brightest flow.
This boundary will be called later on as the bifurcation curve.

We  choose $M_0/M= 5.95496\times 10^{-7}$, where $M_0$ is the Solar mass. 
In the standard gravitational  units $G=c=1$ and  in the scaling $M=1$ one gets 
$\kappa = 2.1326762 \times 10^{21} \left( M_0/M\right)$, that is
$\kappa =1.27\times 10^{15}$. The size of the system is $R_\infty =10^6$. The
speed of sound is given in successive runs by $a^2_\infty =4\times 10^{-3}, 4\times 10^{-4}, 
4\times 10^{-5} $.  The Eddington luminosity reads $L_E=4\pi M/\kappa = 9.9847 \times 10^{-15}$.

\begin{figure}[h]
\includegraphics[width=\linewidth]{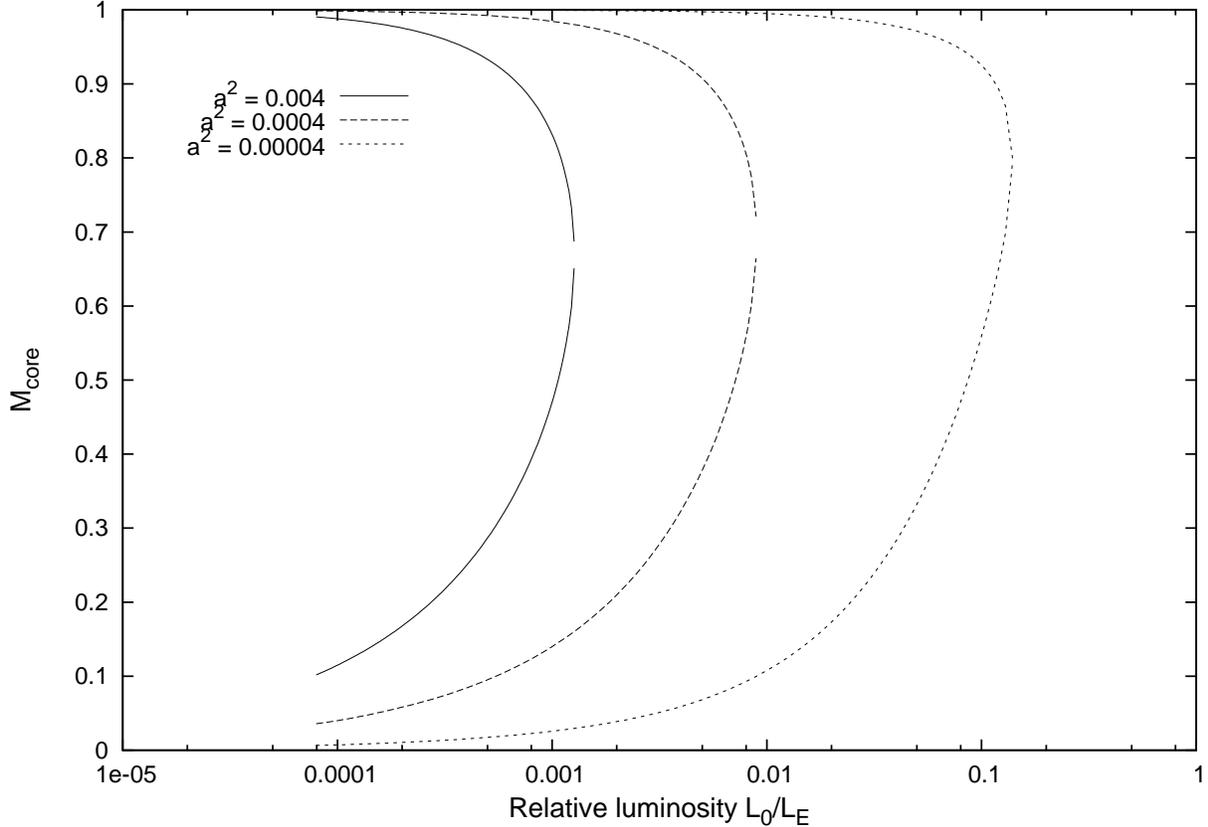}
\caption{\label{fig:1}Small binding energy, $\alpha = 0.0025$. Three asymptotic values
of the speed of sound, $a^2_\infty =0.004, 0.0004, 0.00004$. Two bifurcation branches  
  encompass the set of subsonic flows. The abscissa shows the
luminosity and the ordinate shows the mass of the compact core.}
\end{figure}

\begin{figure}[h]
\includegraphics[width=\linewidth ]{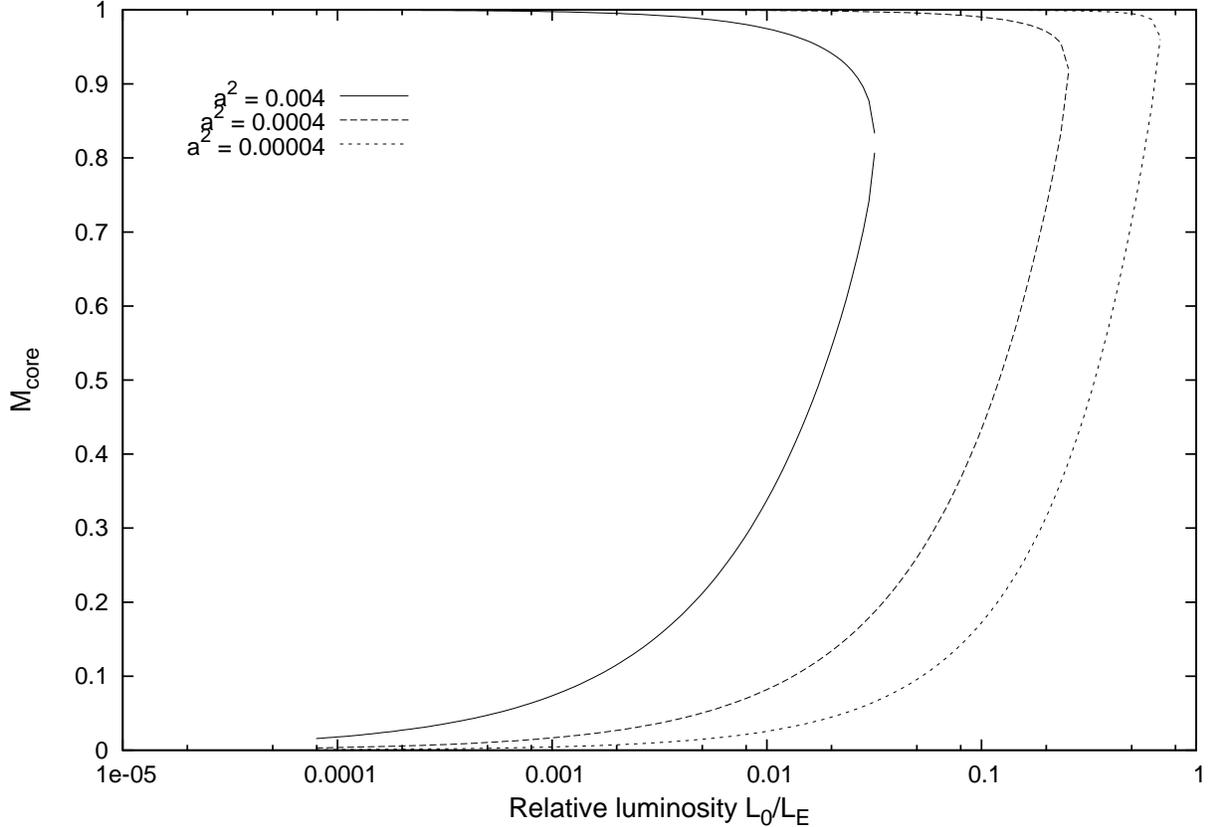}
\caption{\label{fig:2}  $\alpha = 0.5$, $a^2_\infty =0.004, 0.0004, 0.00004$.
  The axes are as in Fig.1.}
\end{figure}

\begin{figure}[h]
\includegraphics[width=\linewidth]{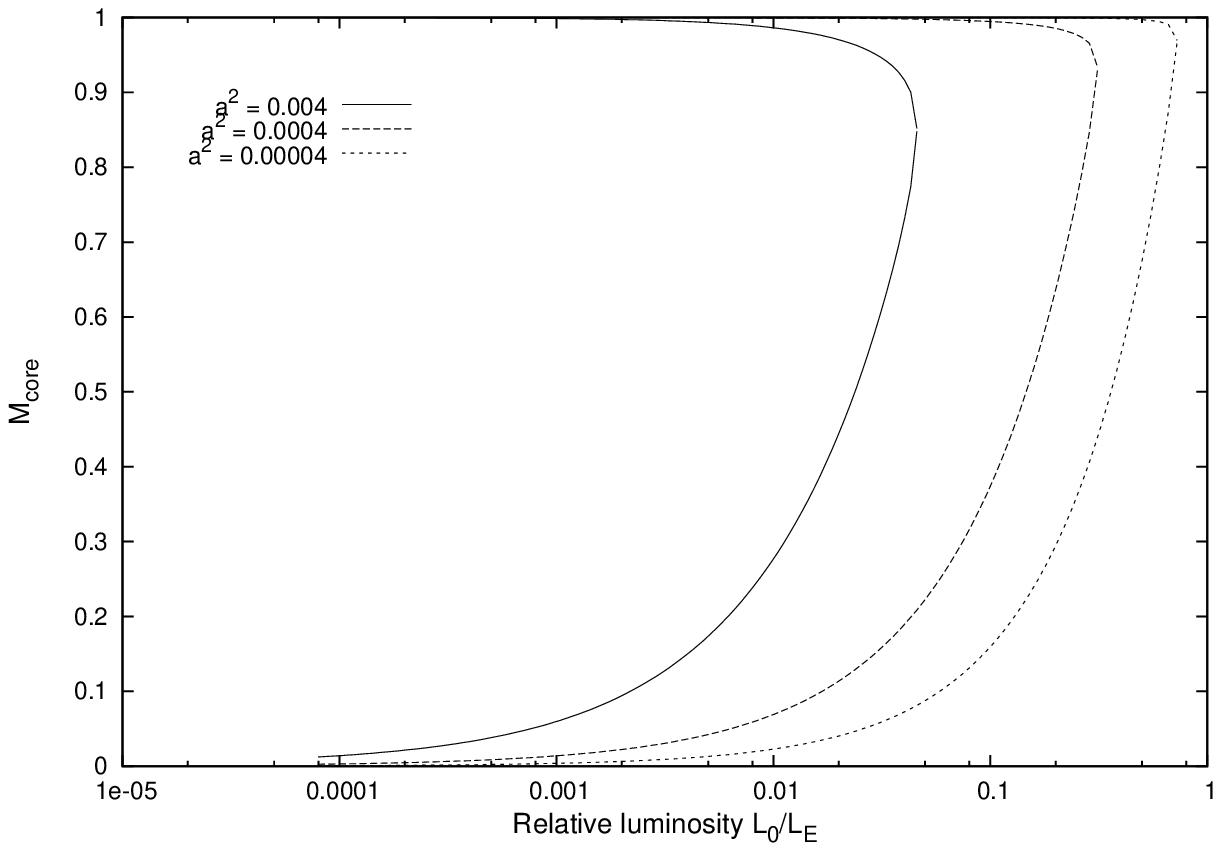}
\caption{\label{fig:3}High binding energy, $\alpha = 0.9,
a^2_\infty =0.004, 0.0004, 0.00004$.  The axes are as in Fig.1.}
\end{figure}

Figures \ref{fig:1}--\ref{fig:3} show accreting solutions on the  
luminosity-(mass of the central core) diagram for 
$\alpha =25\times 10^{-4}, 0.5,0.9$, respectively. Each figure  depicts solution sets for three 
different values of the asymptotic speed of sound $a^2_\infty $.
 For  small $\alpha$   there can exist two accreting 
solutions possessing sonic points, with asymptotic densities $\rho_{0 \infty 1}$
and $\rho_{0 \infty 2}$.  Subsonic flows  then   exist for  each $\rho_{0\infty }
\in \left(\rho_{0 \infty 1 }, \rho_{0 \infty 2} \right)$. Subsonic solutions are not
specified uniquely for given boundary data  but the
length of the interval of allowed values of the asymptotic baryonic density
$\rho_{0\infty }$ becomes shorter with the increase of $L_0$.
 
For larger     $L_0$ and/or $\alpha $  the   bifurcation curve  
can consist either  of subsonic or supersonic flows and its interior consists
of subsonic solutions \cite{Karkowski2008}. The brightest system 
 coincides, as before,  with the bifurcation point and it is unique. The luminosity of the bifurcation 
point increases with the decrease of the asymptotic speed of sound
and it goes up with the increase of $\alpha $. Its
gas abundance  depends  both on luminosity and redshift.

The gas abundance for 
$\alpha =25\times 10^{-4}$ decreases from almost $1/3$ at $a^2_\infty =0.004$
to $1/5$ at $a^2_\infty =0.00004$, as shown on Fig. 1. This 
value of $\alpha $ implies $2M(R_0)/R_0\approx 0.005$.   
Interestingly, the abundance $1/3$ can be shown analytically  to characterize those 
general relativistic polytropic flows without radiation that  maximize the accretion rate  
\cite{PRD2006} and low luminosity newtonian sources \cite{AA}.    
The case of $\alpha = 0.5$ corresponds to a very compact central object with 
$2M(R_0)/R_0\approx 0.75$, close to the Buchdahl limit
 \cite{Buchdahl}. Fig. 2 demonstrates that  gas contribution equals about $0.16$
for  $a^2_\infty =0.004$ and  goes down to    $0.04$ for $a^2_\infty =0.00004$. 
When  $\alpha = 0.9$, then   $2M(R_0)/R_0\approx 0.99$ at the surface of the compact
central object, well beyond  the Buchdahl limit. Only exotic matter violating standard 
energy conditions can be responsible for such compact bodies \cite{Mazur}. We find from
Fig. 3 that  gas contribution to the mass changes  from  $1/8$
($a^2_\infty =0.004$) to $1/30$ ($a^2_\infty =0.00004$). 

It is clear that a reformulation of the problem   would allow
one to estimate the mass of  an isolated  system, assuming that the mass of 
the central core is known. This can be possibly applied to Thorne-\.Zytkow stars. 
  
One can show that in models with test fluids the gas density is bounded from below.
In particular, in  a Shakura model and for general relativistic 
systems with low luminosity and redshift,  the bound  is provided by a supersonic
flow \cite{Karkowski2008}. The full general-relativistic analysis reveals a new  
qualitative effect.   Namely in steady (sub-or supersonic) accretion solutions the  
gas abundance is bounded both from below and from above by bounds that depend on 
the redshift and luminosity.  This is a clear demonstration of the importance 
of backreaction in accretion processes.  

The triple of observables $\alpha, L_0, a^2_\infty $ 
does not specify accretion completely, but for high luminosities the remaining 
freedom (in choosing the asymptotic baryonic mass density) is severely restricted 
and the brightest flow is unique.  It is interesting to note that the concept of Eddington
luminosity still  applies  -- in the light of our data -- in the general relativistic case.
In all studied examples we have $L_0 < L_E$; this is supported also by an analytic argument,
discussed above. Now $L_E= 4\pi M/\kappa $; it is the 
global mass rather than the mass of the compact core, that enters the expression for the
Eddington luminosity.

Numerical data show that gas can be abundant in quasi-stationary accreting systems. 
Brightest systems    can possess even 33\% of gas for small redshifts  and still more
than 10\% of gas for $\alpha =0.9$. It is an open problem
whether this conclusion is true in nonspherical steady flows.

\section*{Acknowledgments.}
This paper has been partially supported by the MNII grant 1PO3B 01229.
Zdobys\l aw \'Swierczy\'nski thanks the Pedagogical University for the research
grant. Krzysztof Roszkowski thanks the Foundation for Polish Science for
financial support.

\end{document}